\documentclass{article}
\usepackage{makeidx}
\usepackage{amsfonts}
\usepackage{amsmath}

\newtheorem{theorem}{Theorem}

\newtheorem{corollary}[theorem]{Corollary}

\newtheorem{definition}[theorem]{Definition}

\newtheorem{proposition}[theorem]{Proposition}
\newtheorem{remark}[theorem]{Remark}

\newenvironment{proof}[1][Proof]{\noindent\textbf{#1.} }{\ \rule{0.5em}{0.5em}}
\input{tcilatex}

\begin{document}

\title{Complete families of solutions for the Dirac equation: an application
of bicomplex pseudoanalytic function theory and transmutation operators}
\author{Hugo M. Campos, Vladislav V. Kravchenko and Luis M. M\'{e}ndez \\
{\small Department of Mathematics, CINVESTAV del IPN, Unidad Queretaro, }\\
{\small Libramiento Norponiente No. 2000, Fracc. Real de Juriquilla,
Queretaro, }\\
{\small Qro. C.P. 76230 MEXICO e-mail: vkravchenko@qro.cinvestav.mx\thanks{%
Research was supported by CONACYT, Mexico. Hugo Campos additionally
acknowledges the support by FCT, Portugal.}}}
\maketitle

\begin{abstract}
The Dirac equation with a scalar and an electromagnetic potentials is
considered. In the time-harmonic case and when all the involved functions
depend only on two spatial variables it reduces to a pair of decoupled
bicomplex Vekua-type equations \cite{KrAntonio}. Using the technique
developed for complex Vekua equations a system of exact solutions for the
bicomplex equation is conctructed under additional conditions, in particular
when the electromagnetic potential is absent and the scalar potential is a
function of one Cartesian variable. Introducing a transmutation operator
relating the involved bicomplex Vekua equation with the Cauchy-Riemann
equation we prove the expansion and the Runge approximation theorems
corresponding to the constructed family of solutions.
\end{abstract}

\section{Introduction}

The Dirac system with a scalar and an electromagnetic potentials is
considered. In \cite{KrAntonio} (see also \cite{APFT}) it was shown that in
the time-harmonic case and when the whole model is independent of one of the
spatial variables the system reduces to a pair of decoupled Vekua-type
equations which differ from the classical Vekua equations considered in the
theory of generalized analytic or pseudoanalytic functions \cite{Berskniga}, 
\cite{APFT}, \cite{Vekua} by the fact that they are bicomplex. In \cite%
{KrAntonio} using this reduction as well as a procedure introduced by L.
Bers, for an arbitrary scalar potential depending on one Cartesian variable
an infinite family of solutions of the Dirac system was constructed.
Nevertheless the completeness of this family in the linear space of all
solutions was not proved due to the lack of some fundamental results in the
theory of bicomplex Vekua equations such as the similarity principle and
many other.

The constructed family of solutions is a system of formal powers
generalizing those introduced by L. Bers onto the bicomplex situation.
Meanwhile in the classical complex case there is a well developed theory of
formal powers with the Runge-type approximation theorem and other related
results (see \cite{CCK} and references therein), in the bicomplex case up to
now no such result was available even for simplest examples.

In the present work we prove the completeness of the family of solutions
obtained in \cite{KrAntonio} by using so-called transmutation operators and
some recent results on their mapping properties \cite{CKT}. The notion and
the name of the transmutation operator appeared in the work of J. Delsarte 
\cite{Delsarte1}, \cite{Delsarte2} and later developed in \cite%
{DelsarteLions}, \cite{Lions} and many other publications \cite{Gilbert}, 
\cite{Carroll}, \cite{LevInverse}, \cite{Marchenko}, \cite{Sitnik}, \cite%
{Trimeche}. Combining the results from \cite{CKT} on mapping properties of
the transmutation operators with the results from \cite{KT} on the
construction of a transmutation operator for the Darboux transformed Schr%
\"{o}dinger equation we obtain transmutation operators which relate the
bicomplex Vekua equations arising from the Dirac system with the
Cauchy-Riemann equation. Using this result we prove that the bicomplex
pseudoanalytic formal powers are the result of application of a
corresponding transmutation operator to the usual powers of the complex
variable $z$. This together with the boundedness of the transmutation
operator and of its inverse allows us to prove the expansion and the Runge
approximation theorems for solutions of the considered bicomplex Vekua
equations.

\bigskip

\section{The Dirac system and bicomplex pseudoanalytic functions}

Consider the Dirac operator 
\index{Dirac operator}with a scalar and an electromagnetic potentials%
\begin{equation*}
\mathbb{D}=\gamma _{0}\partial _{t}+\sum_{k=1}^{3}\gamma _{k}\partial
_{k}+i\left( m+p_{el}\gamma _{0}+\sum_{k=1}^{3}A_{k}\gamma _{k}+p_{sc}\right)
\end{equation*}%
where $\gamma _{j},$ $j=0,1,2,3$ are usual $\gamma $-matrices (see, e.g., 
\cite{BD}, \cite{Thaller}) 
\begin{equation*}
\gamma _{0}:=\left( 
\begin{array}{rrrr}
1 & 0 & 0 & 0 \\ 
0 & 1 & 0 & 0 \\ 
0 & 0 & -1 & 0 \\ 
0 & 0 & 0 & -1%
\end{array}%
\right) ,\hskip48pt\gamma _{1}:=\left( 
\begin{array}{rrrr}
0 & 0 & 0 & -1 \\ 
0 & 0 & -1 & 0 \\ 
0 & 1 & 0 & 0 \\ 
1 & 0 & 0 & 0%
\end{array}%
\right) ,
\end{equation*}%
\begin{equation*}
\gamma _{2}:=\left( 
\begin{array}{rrrr}
0 & 0 & 0 & i \\ 
0 & 0 & -i & 0 \\ 
0 & -i & 0 & 0 \\ 
i & 0 & 0 & 0%
\end{array}%
\right) ,\hskip48pt\gamma _{3}:=\left( 
\begin{array}{rrrr}
0 & 0 & -1 & 0 \\ 
0 & 0 & 0 & 1 \\ 
1 & 0 & 0 & 0 \\ 
0 & -1 & 0 & 0%
\end{array}%
\right) ,
\end{equation*}%
$m\in \mathbb{R}$, $p_{el}$, $A_{k}$ and $p_{sc}$ are real valued functions.

We will denote the algebra of biquaternions or complex quaternions by $%
\mathbb{H(C)}$ with the standard basic quaternionic units denoted by $%
e_{0}=1 $, $e_{1},e_{2}$ and $e_{3}$. The complex imaginary unit is denoted
by $i$ as usual. The set of purely vectorial quaternions $q=\mathbf{q}$ is
identified with the set of three-dimensional vectors.

The quaternionic conjugation of a biquaternion $q=q_{0}+\mathbf{q}$ will be
denoted as $%
\overline{q}=q_{0}-\mathbf{q}$. Sometimes the following notation for the
operator of multiplication from the right-hand side will be used $%
M^{p}q=q\cdot p$.

The main quaternionic differential operator introduced by Hamilton himself
and sometimes called the Moisil-Theodoresco operator is defined on
continuously differentiable biquaternion-valued functions of the real
variables $x_{1}$, $x_{2}$ and $x_{3}$ according to the rule%
\begin{equation*}
Dq=\sum_{k=1}^{3}e_{k}\partial _{k}q,
\end{equation*}%
where $\partial _{k}=\frac{\partial }{\partial x_{k}}$.

In \cite{Krbag} (see also \cite{CK2003}, \cite{AQA}, \cite{KSbook}) a simple
invertible matrix transformation was obtained which allows one to rewrite
the classical Dirac equation in biquaternionic terms. Namely, the Dirac
operator $\mathbb{D}$ is equivalent to the biquaternionic operator 
\begin{equation*}
R=D-\partial _{t}M^{e_{1}}+\mathbf{a}+M^{-i(\widetilde{p}_{el}e_{1}-i(%
\widetilde{p}_{sc}+m)e_{2})}
\end{equation*}%
where $\mathbf{a}=i(\widetilde{A}_{1}e_{1}+\widetilde{A}_{2}e_{2}-\widetilde{%
A}_{3}e_{3})$ and the notation \ \textquotedblleft $\widetilde{\cdot }$%
\textquotedblright\ $\ $means the reflection with respect to $x_{3}$, $%
\widetilde{f}:=f(t,x_{1},x_{2},-x_{3})$. Note that in the absence of the
electromagnetic potential the operator $R$ becomes real quaternionic which
is an important property (see \cite{KrRam}).

In what follows we assume that potentials are time-independent and consider
solutions with a fixed energy: $\Phi (t,\mathbf{x})=\Phi _{\omega }(\mathbf{x%
})e^{i\omega t}$. The equation for $\Phi _{\omega }$ has the form%
\begin{equation}
\mathbb{D}_{\omega }\Phi _{\omega }=0\qquad \text{in }\widehat{G}
\label{DiracOmega}
\end{equation}%
where $\widehat{G}$ is a domain in $\mathbb{R}^{3}$,%
\begin{equation*}
\mathbb{D}_{\omega }=i\omega \gamma _{0}+\sum_{k=1}^{3}\gamma _{k}\partial
_{k}+i\left( m+p_{el}\gamma _{0}+\sum_{k=1}^{3}A_{k}\gamma
_{k}+p_{sc}\right) .
\end{equation*}%
Under the mentioned above matrix transformation the operator $\mathbb{D}%
_{\omega }$ turns into its biquaternionic counterpart 
\begin{equation*}
R_{\omega }=D+\mathbf{a}+M^{\mathbf{b}}
\end{equation*}%
with $\mathbf{b}=-i((\widetilde{p}_{el}+\omega )e_{1}-i(\widetilde{p}%
_{sc}+m)e_{2})$. Thus, equation (\ref{DiracOmega}) turns into the complex
quaternionic equation 
\begin{equation}
R_{\omega }q=0  \label{ROmega}
\end{equation}%
where $q$ is a complex quaternion valued function. In what follows we study
this equation.

Let us introduce the following notation. For any biquaternion $q$ we denote
by $Q_{1}$ and $Q_{2}$ its bicomplex components%
\index{bicomplex number}:%
\begin{equation*}
Q_{1}=q_{0}+q_{3}e_{3}\qquad 
\text{and\qquad }Q_{2}=q_{2}-q_{1}e_{3}.
\end{equation*}%
Then $q$ can be represented as follows $q=Q_{1}+Q_{2}e_{2}$. For the
operator $D$ we have $D=D_{1}+D_{2}e_{2}$ with $D_{1}=e_{3}\partial _{3}$
and $D_{2}=\partial _{2}-\partial _{1}e_{3}$. Notice that $\mathbf{b}=Be_{2}$
with $B=-(\widetilde{p}_{sc}+m)+i(\widetilde{p}_{el}+\omega )e_{3}$, $%
\mathbf{a}=A_{1}+A_{2}e_{2}$ with $A_{1}=a_{3}e_{3}$ and $%
A_{2}=a_{2}-a_{1}e_{3}$.

We obtain that equation (\ref{ROmega}) is equivalent to the system%
\begin{equation}
D_{1}Q_{1}-D_{2}\overline{Q}_{2}+A_{1}Q_{1}-A_{2}\overline{Q}_{2}-\overline {%
B}Q_{2}=0,  \label{Diracsys1}
\end{equation}%
\begin{equation}
D_{2}\overline{Q}_{1}+D_{1}Q_{2}+A_{2}\overline{Q}_{1}+A_{1}Q_{2}+BQ_{1}=0,
\label{Diracsys2}
\end{equation}
where $Q_{1}$ and $Q_{2}$ are bicomplex components of $q$. We stress \ that
the system (\ref{Diracsys1}), (\ref{Diracsys2}) is equivalent to the Dirac
equation in $\gamma$-matrices (\ref{DiracOmega}).

Let us suppose all fields in our model to be independent of $x_{3}$, and $%
A_{1}=a_{3}e_{3}\equiv 0$. Then the system (\ref{Diracsys1}), (\ref%
{Diracsys2}) decouples, and we obtain two separate bicomplex equations \cite%
{KrAntonio}, \cite{APFT} 
\begin{equation*}
\overline{D}_{2}Q_{2}=-\overline{A}_{2}Q_{2}-B\overline{Q}_{2},\quad \text{%
and}\quad \overline{D}_{2}Q_{1}=-\overline{A}_{2}Q_{1}-\overline{B}\overline{%
Q}_{1}.
\end{equation*}%
Denote $\overline{\partial }=\overline{D}_{2}$, $a=-\overline{A}_{2}$, $b=-B$%
, $w=Q_{2}$, $W=Q_{1}$, $z=x+y\mathbf{k}$, where $x=x_{2}$, $y=x_{1}$ and
for convenience we denote $\mathbf{k}=e_{3}$. Then we reduce the Dirac
equation with electromagnetic and scalar potentials independent of $x_{3}$
to a pair of Vekua-type equations%
\begin{equation}
\overline{\partial }w=aw+b\overline{w}  \label{Vekua1D}
\end{equation}%
and 
\begin{equation}
\overline{\partial }W=aW+\overline{bW}.  \label{Vekua2}
\end{equation}

\section{Some definitions and results from bicomplex pseudoanalytic function
theory}

\begin{definition}
We consider $\mathbb{B}$-valued functions of two real variables $x$ and $y$.
Denote $\overline{\partial }=\frac{1}{2}(\frac{\partial }{\partial x}+%
\mathbf{k}\frac{\partial }{\partial y})$ and $\partial =\frac{1}{2}(\frac{%
\partial }{\partial x}-\mathbf{k}\frac{\partial }{\partial y})$. An equation
of the form 
\begin{equation}
\overline{\partial }w=aw+b\overline{w},  \label{Vekuabic}
\end{equation}%
where $w$, $a$ and $b$ are $\mathbb{B}$-valued functions is called a
bicomplex Vekua equation. When all the involved functions have their values
in $\mathbb{C}_{\mathbf{k}}$ only, equation (\ref{Vekuabic}) becomes the
well known complex Vekua equation (see \cite{APFT}, \cite{Vekua}). We will
assume that $w\in C^{1}(\Omega )$ where $\Omega \subset \mathbb{R}^{2}$ is
an open domain and $a$, $b$ are H\"{o}lder continuous in $\Omega $.
\end{definition}

When $a\equiv0$ and $b=\frac{\overline{\partial}\phi}{\phi}$ where $\phi:%
\overline{\Omega}\rightarrow\mathbb{C}_{i}$ possesses H\"{o}lder continuous
partial derivatives in $\Omega$ and $\phi(x,y)\neq0$, $\forall(x,y)\in%
\overline{\Omega}$ we will say that the bicomplex Vekua equation 
\begin{equation}
\overline{\partial}w=\frac{\overline{\partial}\phi}{\phi}\overline {w}
\label{Vekuamain}
\end{equation}
is a Vekua equation of the main type or the main Vekua equation.

For classical complex Vekua equations Bers introduced \cite{Berskniga} the
notions of a generating pair, generating sequence, formal powers and Taylor
series in formal powers. As was shown in \cite{KrAntonio}, \cite{APFT} the
definition of these notions can be extended onto the bicomplex situation.
Here we briefly recall the main definitions.

\begin{definition}
A pair of $\mathbb{B}$-valued functions $F$ and $G$ possessing H\"{o}lder
continuous partial derivatives in $\Omega$ with respect to the real
variables $x$ and $y$ is said to be a generating pair if it satisfies the
inequality%
\begin{equation}
\func{Vec}(\overline{F}G)\neq0\qquad\text{in }\Omega.  \label{condGenPair}
\end{equation}
\end{definition}

Condition (\ref{condGenPair}) implies that every bicomplex function $w$
defined in a subdomain of $\Omega$ admits the unique representation $w=\phi
F+\psi G$ where the functions $\phi$ and $\psi$ are scalar ($\mathbb{C}_{i}$%
-valued).

\begin{remark}
\label{RemBanalytic}When $F\equiv 1$ and $G\equiv \mathbf{k}$ the
corresponding bicomplex Vekua equation is 
\begin{equation}
\overline{\partial }w=0,  \label{C-Rbic}
\end{equation}%
and its study in fact reduces to the complex analytic function theory \cite%
{CKT}. Indeed, consider the following pair of idempotents $\mathbf{P}^{+}=%
\frac{1}{2}(1+i\mathbf{k})$ and $\mathbf{P}^{-}=\frac{1}{2}(1-i\mathbf{k})$ (%
$\left( \mathbf{P}^{\pm }\right) ^{2}=\mathbf{P}^{\pm }$). Then the
functions $\mathbf{P}^{+}w$ and $\mathbf{P}^{-}w$ are necessarily
antiholomorphic and holomorphic respectively. Indeed, application of $%
\mathbf{P}^{+}$ and $\mathbf{P}^{-}$ to (\ref{C-Rbic}) gives us 
\begin{equation}
\partial _{z}\mathbf{P}^{+}w=0\quad \text{and}\quad \partial _{\overline{z}}%
\mathbf{P}^{-}w=0  \label{dP}
\end{equation}%
where $\partial _{z}=\frac{1}{2}(\frac{\partial }{\partial x}-i\frac{%
\partial }{\partial y})$ and $\partial _{\overline{z}}=\frac{1}{2}(\frac{%
\partial }{\partial x}+i\frac{\partial }{\partial y})$. Moreover, $\mathbf{P}%
^{+}w=\mathbf{P}^{+}(u+jv)=\mathbf{P}^{+}(u-iv)$ and $\mathbf{P}^{-}w=%
\mathbf{P}^{-}(u+iv)$. Due to (\ref{dP}) the scalar functions $w^{+}:=u-iv$
and $w^{-}:=u+iv$ are antiholomorphic and holomorphic respectively. We
stress that $w^{+}$ is not necessarily a complex conjugate of $w^{-}$ ($u$
and $v$ are $\mathbb{C}_{i}$-valued).

Let us notice that due to the equivalence of (\ref{C-Rbic}) and (\ref{dP})
we have that a bicomplex solution $w$ of (\ref{C-Rbic}) admits a convergent
Taylor series $w(z)=\sum_{n=0}^{\infty }a_{n}z^{n}$ if and only if the
series $\sum_{n=0}^{\infty }a_{n}^{+}\left( z^{+}\right) ^{n}$ and $%
\sum_{n=0}^{\infty }a_{n}^{-}\left( z^{-}\right) ^{n}$ corresponding to $%
w^{+}$ and $w^{-}$ respectively converge (here $a_{n}^{\pm }$ and $z^{\pm }$
are scalars, $a_{n}^{\pm }=Sc\left( a_{n}\right) \mp iVec\left( a_{n}\right) 
$ and $z^{\pm }=x\mp iy$). In particular, the radius of convergence of the
power series $\sum_{n=0}^{\infty }a_{n}z^{n}$ has the form $R=\min \left\{
R_{+},R_{-}\right\} $ where $1/R_{\pm }=\lim \sup_{n\rightarrow \infty
}\left\vert a_{n}^{\pm }\right\vert ^{1/n}$.
\end{remark}

Assume that $(F,G)$ is a generating pair in a domain $\Omega $.

\begin{definition}
Let the $\mathbb{B}$-valued function $w$ be defined in a neighborhood of $%
z_{0}\in \Omega \subset \mathbb{C}_{\mathbf{k}}$. In a complete analogy with
the complex case we say that at $z_{0}$ the function $w$ possesses the $%
(F,G) $-derivative%
\index{(F,G)-derivative} $\overset{\cdot }{w}(z_{0})$ if the (finite) limit 
\begin{equation}
\overset{\cdot }{w}(z_{0})=\lim_{z\rightarrow z_{0}}%
\frac{w(z)-\lambda _{0}F(z)-\mu _{0}G(z)}{z-z_{0}}  \label{derivative_def}
\end{equation}%
exists where $\lambda _{0}$ and $\mu _{0}$ are the unique scalar constants
such that $w(z_{0})=\lambda _{0}F(z_{0})+\mu _{0}G(z_{0})$.
\end{definition}

Similarly to the complex case (see, e.g., \cite[Chapter 2]{APFT}) it is easy
to show that if $\overset{\cdot}{w}(z_{0})$ exists then at $z_{0}$, $%
\overline{\partial}w$ and $\partial w$ exist and equations 
\begin{equation}
\overline{\partial}w=a_{(F,G)}w+b_{(F,G)}\overline{w}  \label{Vekua_equation}
\end{equation}
and 
\begin{equation}
\overset{\cdot}{w}=\partial w-A_{(F,G)}w-B_{(F,G)}\overline{w}
\label{derivative_with_characteristic}
\end{equation}
hold, where $a_{(F,G)}$, $b_{(F,G)}$, $A_{(F,G)}$ and $B_{(F,G)}$ are the 
\emph{characteristic coefficients}%
\index{characteristic coefficients} of the pair $(F,G)$ defined by the
formulas 
\begin{equation*}
a_{(F,G)}=-%
\frac{\overline{F}\,\overline{\partial}G-\overline{G}\,\overline{\partial}F}{%
F\overline{G}-\overline{F}G},\qquad b_{(F,G)}=\frac{F\,\overline{\partial}%
G-G\,\overline{\partial}F}{F\overline {G}-\overline{F}G},
\end{equation*}

\begin{equation*}
A_{(F,G)}=-\frac{\overline{F}\,\partial G-\overline{G}\,\partial F}{F%
\overline{G}-\overline{F}G},\qquad B_{(F,G)}=\frac{F\,\partial G-G\,\partial
F}{F\overline{G}-\overline{F}G}.
\end{equation*}%
Note that $F\overline{G}-\overline{F}G=-2\mathbf{k}\func{Vec}(\overline{F}%
G)\neq 0$.

If $\overline{\partial}w$ and $\partial w$ exist and are continuous in some
neighborhood of $z_{0}$, and if (\ref{Vekua_equation}) holds at $z_{0}$,
then $\overset{\cdot}{w}(z_{0})$ exists, and (\ref%
{derivative_with_characteristic}) holds. Let us notice that $F$ and $G$
possess $(F,G)$-derivatives, $\overset{\cdot}{F}\equiv\overset{\cdot}{G}%
\equiv0$ and the following equalities are valid which determine the
characteristic coefficients uniquely%
\begin{equation*}
\overline{\partial}F=a_{(F,G)}F+b_{(F,G)}\overline{F},\quad\overline{%
\partial }G=a_{(F,G)}G+b_{(F,G)}\overline{G},
\end{equation*}%
\begin{equation*}
\partial F=A_{(F,G)}F+B_{(F,G)}\overline{F},\quad\partial
G=A_{(F,G)}G+B_{(F,G)}\overline{G}.
\end{equation*}
If the $(F,G)$-derivative of a $\mathbb{B}$-valued function $w=\phi F+\psi G$
(where the functions $\phi$ and $\psi$ are scalar) exists, besides the form (%
\ref{derivative_with_characteristic}) it can also be written as follows $%
\overset{\cdot}{w}=\partial\phi\,F+\partial\psi\,G$.

\begin{definition}
\label{DefSuccessor_bi}Let $(F,G)$ and $(F_{1},G_{1})$ -- be two generating
pairs in $\Omega$. $(F_{1},G_{1})$ is called \ successor of $(F,G)$ and $%
(F,G)$ is called predecessor of $(F_{1},G_{1})$ if%
\begin{equation*}
a_{(F_{1},G_{1})}=a_{(F,G)}\qquad\text{and}\qquad
b_{(F_{1},G_{1})}=-B_{(F,G)}\text{.}
\end{equation*}
\end{definition}

By analogy with the complex case we have the following statement.

\begin{theorem}
\label{ThBersDer_bi}Let $w$ be a bicomplex $(F,G)$-pseudoanalytic function
and let $(F_{1},G_{1})$ be a successor of $(F,G)$. Then $\overset{\cdot}{w}$
is a bicomplex $(F_{1},G_{1})$-pseudoanalytic function.
\end{theorem}

\begin{definition}
\label{DefAdjoint_bi}Let $(F,G)$ be a generating pair. Its adjoint
generating pair $(F,G)^{\ast}=(F^{\ast},G^{\ast})$ is defined by the formulas%
\begin{equation*}
F^{\ast}=-\frac{2\overline{F}}{F\overline{G}-\overline{F}G},\qquad G^{\ast }=%
\frac{2\overline{G}}{F\overline{G}-\overline{F}G}.
\end{equation*}
\end{definition}

The $(F,G)$-integral is defined as follows 
\begin{equation*}
\int_{\Gamma}Wd_{(F,G)}z=\frac{1}{2}\left( F(z_{1})\func{Sc}%
\int_{\Gamma}G^{\ast}Wdz+G(z_{1})\func{Sc}\int_{\Gamma}F^{\ast }Wdz\right)
\end{equation*}
where $\Gamma$ is a rectifiable curve leading from $z_{0}$ to $z_{1}$.

If $W=\phi F+\psi G$ is a bicomplex $(F,G)$-pseudoanalytic function where $%
\phi $ and $\psi $ are complex valued functions then 
\begin{equation}
\int_{z_{0}}^{z}\overset{\cdot }{W}d_{(F,G)}z=W(z)-\phi (z_{0})F(z)-\psi
(z_{0})G(z),  \label{FGAntD}
\end{equation}%
and this integral is path-independent and represents the $(F,G)$%
-antiderivative of $\overset{\cdot }{W}$.

\begin{definition}
\label{DefSeq_bi}A sequence of generating pairs $\left\{
(F_{m},G_{m})\right\} $, $m=0,\pm1,\pm2,\ldots$ , is called a generating
sequence if $(F_{m+1},G_{m+1})$ is a successor of $(F_{m},G_{m})$. If $%
(F_{0},G_{0})=(F,G)$, we say that $(F,G)$ is embedded in $\left\{
(F_{m},G_{m})\right\} $.
\end{definition}

Let $W$ be a bicomplex $(F,G)$-pseudoanalytic function. Using a generating
sequence in which $(F,G)$ is embedded we can define the higher derivatives
of $W$ by the recursion formula%
\begin{equation*}
W^{[0]}=W;\qquad W^{[m+1]}=\frac{d_{(F_{m},G_{m})}W^{[m]}}{dz},\quad
m=1,2,\ldots\text{.}
\end{equation*}

\begin{definition}
\label{DefFormalPower_bi}The formal power $Z_{m}^{(0)}(a,z_{0};z)$ with
center at $z_{0}\in\Omega$, coefficient $a$ and exponent $0$ is defined as
the linear combination of the generators $F_{m}$, $G_{m}$ with scalar
constant coefficients $\lambda$, $\mu$ chosen so that $\lambda
F_{m}(z_{0})+\mu G_{m}(z_{0})=a$. The formal powers with exponents $%
n=0,1,2,\ldots$ are defined by the recursion formula%
\begin{equation}
Z_{m}^{(n+1)}(a,z_{0};z)=(n+1)\int_{z_{0}}^{z}Z_{m+1}^{(n)}(a,z_{0};%
\zeta)d_{(F_{m},G_{m})}\zeta.  \label{recformulaD}
\end{equation}
\end{definition}

This definition implies the following properties.

\begin{enumerate}
\item $Z_{m}^{(n)}(a,z_{0};z)$ is an $(F_{m},G_{m})$-pseudoanalytic function
of $z$.

\item If $a^{\prime }$ and $a^{\prime \prime }$ are scalar constants, then 
\begin{equation*}
Z_{m}^{(n)}(a^{\prime }+\mathbf{k}a^{\prime \prime },z_{0};z)=a^{\prime
}Z_{m}^{(n)}(1,z_{0};z)+a^{\prime \prime }Z_{m}^{(n)}(\mathbf{k},z_{0};z).
\end{equation*}

\item The formal powers satisfy the differential relations%
\begin{equation*}
\frac{d_{(F_{m},G_{m})}Z_{m}^{(n)}(a,z_{0};z)}{dz}%
=nZ_{m+1}^{(n-1)}(a,z_{0};z).
\end{equation*}

\item The asymptotic formulas 
\begin{equation*}
Z_{m}^{(n)}(a,z_{0};z)\sim a(z-z_{0})^{n},\quad z\rightarrow z_{0}
\end{equation*}
hold.
\end{enumerate}

The case of the main bicomplex Vekua equation is of a special interest also
due to the following relation with the stationary Schr\"{o}dinger equation.

\begin{theorem}
\cite{KrJPhys06} Let $W=W_{1}+\mathbf{k}W_{2}$ be a solution of the main
bicomplex Vekua equation 
\begin{equation}
\overline{\partial }W=\frac{\overline{\partial }\phi }{\phi }\overline{W}%
\quad \text{in }\Omega  \label{mainV}
\end{equation}%
where $W_{1}=\func{Sc}W$, $W_{2}=\func{Vec}W$ and the $\mathbb{C}_{i}$%
-valued function $\phi $ is a nonvanishing solution of the equation 
\begin{equation}
-\Delta u+q_{1}(x,y)u=0\quad \text{in }\Omega  \label{eqq1}
\end{equation}%
where $q_{1}$ is a continuous $\mathbb{C}_{i}$-valued function. Then $W_{1}$
is a solution of (\ref{eqq1}) in $\Omega $ and $W_{2}$ is a solution of the
associated Schr\"{o}dinger equation 
\begin{equation}
-\Delta v+q_{2}(x,y)v=0\quad \text{in }\Omega  \label{eqq2}
\end{equation}%
where $q_{2}=8\frac{\overline{\partial }\phi \,\partial \phi }{\phi ^{2}}%
-q_{1}$.
\end{theorem}

We need the following notation. Let $w$ be a $\mathbb{B}$-valued function
defined on a simply connected domain $\Omega$ with $w_{1}=\func{Sc}w$ and $%
w_{2}=\func{Vec}w$ such that 
\begin{equation}
\frac{\partial w_{1}}{\partial y}-\frac{\partial w_{2}}{\partial x}%
=0,\quad\forall\,(x,y)\in\Omega,  \label{compcond}
\end{equation}
and let $\Gamma\subset\Omega$ be a rectifiable curve leading from $%
(x_{0},y_{0})$ to $(x,y)$. Then the integral 
\begin{equation*}
\overline{A}w(x,y):=2\left( \int_{\Gamma}w_{1}dx+w_{2}dy\right)
\end{equation*}
is path-independent, and all $\mathbb{C}_{i}$-valued solutions $\varphi$ of
the equation $\overline{\partial}\varphi=w$ in $\Omega$ have the form $%
\varphi(x,y)=\overline{A}w(x,y)+c$ where $c$ is an arbitrary $\mathbb{C}_{i}$%
-constant. In other words the operator $\overline{A}$ denotes the well known
operation for reconstructing the potential function from its gradient.

\begin{theorem}
\cite{KrJPhys06} Let $W_{1}$ be a $\mathbb{C}_{i}$-valued solution of the
Schr\"{o}dinger equation (\ref{eqq1}) in a simply connected domain $\Omega $%
. Then a $\mathbb{C}_{i}$-valued solution $W_{2}$ of the associated Schr\"{o}%
dinger equation (\ref{eqq2}) such that $W_{1}+\mathbf{k}W_{2}$ is a solution
of (\ref{mainV}) in $\Omega $ can be constructed according to the formula 
\begin{equation*}
W_{2}=\frac{1}{\phi }\overline{A}\left( \mathbf{k}\,\phi ^{2}\,\overline{%
\partial }\left( \frac{W_{1}}{\phi }\right) \right) +\frac{c_{1}}{\phi }
\end{equation*}%
where $c_{1}$ is an arbitrary $\mathbb{C}_{i}$-constant.

Vice versa, given a solution $W_{2}$ of (\ref{eqq2}), the corresponding \
solution $W_{1}$ of (\ref{eqq1}) such that $W_{1}+\mathbf{k}W_{2}$ is a
solution of (\ref{mainV}) has the form 
\begin{equation*}
W_{1}=-\phi \overline{A}\left( \frac{\mathbf{k}}{\phi ^{2}}\,\overline{%
\partial }\left( \phi W_{2}\right) \right) +c_{2}\phi
\end{equation*}%
where $c_{2}$ is an arbitrary $\mathbb{C}_{i}$-constant.
\end{theorem}

As was shown in \cite{KrRecentDevelopments} (see also \cite{APFT}) a
generating sequence can be obtained in a closed form, for example, in the
case when $\phi $ has a separable form $\phi =S(s)T(t)$ where $s$ and $t$
are conjugate harmonic functions and $S$, $T$ are arbitrary twice
continuously differentiable functions. In practical terms this means that
whenever the Schr\"{o}dinger equation (\ref{eqq1}) admits a particular
nonvanishing solution having the form $\phi =f(\xi )\,g(\eta )$ where $(\xi
,\eta )$ is one of the encountered in physics orthogonal coordinate systems
in the plane a generating sequence corresponding to (\ref{mainV}) can be
obtained explicitly \cite[Sect. 4.8]{APFT}. The knowledge of a generating
sequence allows one to construct the formal powers following Definition \ref%
{DefFormalPower_bi}. This construction is a simple algorithm which can be
quite easily and efficiently realized numerically \cite{CCK}, \cite{CKR}.
Moreover, in the case of a complex main Vekua equation which in the
notations admitted in the present paper corresponds to the case of $\phi $
being a real-valued function (then the main bicomplex Vekua equation
decouples into two main complex Vekua equations) the completeness of the
system of formal powers was proved \cite{CCK} in the sense that any
pseudoanalytic in $\Omega $ and H\"{o}lder continuous on $\partial \Omega $
function can be approximated uniformly and arbitrarily closely by a finite
linear combination of the formal powers. The real parts of the complex
pseudoanalytic formal powers represent then a complete system of solutions
of one Schr\"{o}dinger equation meanwhile the imaginary parts give us a
complete system of solutions of the associated Schr\"{o}dinger equation.

\section{Transmutation operators and a complete family of solutions of the
Dirac equation}

Let us consider the following situation $p_{sc}=p(x),$ $p_{el}=0$ and $%
\overrightarrow{A}=0$. Then the Dirac equation is equivalent to the pair of
bicomplex Vekua equations%
\begin{equation}
\overline{\partial }w=b\overline{w}  \label{6}
\end{equation}%
\begin{equation}
\overline{\partial }W=\overline{bW}  \label{7}
\end{equation}%
where $b=p(x)+m-i\omega \mathbf{k}$.

Let $P$ denote an antiderivative of $p$. Consider the function%
\begin{equation*}
\phi (x,y)=e^{P(x)+mx+i\omega y}
\end{equation*}%
Note that $\overline{\partial }\phi /\phi =\overline{b}$. Then if $W$ is a
solution of $\left( \ref{7}\right) $ then the complex valued function $%
W_{1}=Sc(W)$ is a solution of the Schr\"{o}dinger equation 
\begin{equation*}
\left( -\Delta +\nu \right) W_{1}=0,\text{ with }\upsilon (x)=p\text{%
\'{}%
}(x)+\left( p(x)+m\right) ^{2}-\omega ^{2}
\end{equation*}%
and the complex valued function $W_{2}=Vec(W)$ is a solution of the
associated Schr\"{o}dinger equation 
\begin{equation*}
\left( -\Delta +\mu \right) W_{2}=0,\text{ with }\mu (x)=-p\text{%
\'{}%
}(x)+\left( p(x)+m\right) ^{2}-\omega ^{2}
\end{equation*}

On the other hand equation $\left( \ref{7}\right) $ can be written as a main
Vekua equation%
\begin{equation}
\left( \overline{\partial }-\frac{\overline{\partial }\phi }{\phi }C\right)
W=0  \label{8}
\end{equation}%
where 
\begin{equation*}
\phi (x,y)=f(x)g(y)\text{ with }f(x)=e^{P(x)+mx}\text{ \ and \ }%
g(y)=e^{i\omega y}
\end{equation*}%
Notice that $f$ and $g$ are complex valued functions. We assume that their
domains of definitions are finite segments $\left[ -a,a\right] $ and $\left[
-b,b\right] $ respectively. Assuming that $p\in C^{1}\left[ -a,a\right] $ we
obtain that $f$ and $g$ are nonvanishing $C^{2}$-functions. The separable
form of $\phi $ allows us to write down a generating pair associated with
equation (\ref{8}) $\left( F,G\right) =(\phi ,\mathbf{k}/\phi )$ as well as
a generating sequence of the period two embedding this generating pair 
\begin{eqnarray*}
\text{\ }\left( F,G\right) &=&(\phi ,\mathbf{k}/\phi )\text{; \ }\left(
F_{1},G_{1}\right) =\left( \phi /f^{2},\mathbf{k}f^{2}/\phi \right) \text{ \ 
} \\
\text{ \ }\left( F_{2},G_{2}\right) &=&\left( F,G\right) \text{ \ ; \ \ }%
\left( F_{3},G_{3}\right) =\left( F_{1},G_{1}\right) \text{ \ ; \ }....
\end{eqnarray*}%
The corresponding formal powers can be constructed as follows. We consider
the formal powers with the centre in the origin and for simplicity assume
that $f(0)=1$ (for $g$ this is also the case). Define the following systems
of functions $\left\{ \varphi _{k}\right\} _{k=0}^{\infty }$ and $\left\{
\psi _{k}\right\} _{k=0}^{\infty }$ 
\begin{equation}
\varphi _{k}(x)=\QATOPD\{ . {f(x)X^{(k)}(x)\text{, }k\text{ odd}}{f(x)%
\widetilde{X}^{(k)}(x)\text{, }k\text{ even}}  \label{9}
\end{equation}%
where 
\begin{eqnarray*}
X^{(0)}(x) &=&\widetilde{X}^{(0)}(x)=1 \\
\text{\ }X^{(n)}(x) &=&n\int_{o}^{x}X^{(n-1)}(\rho )\left[ f^{2}(\rho )%
\right] ^{(-1)^{n}}d\rho \text{ } \\
\text{\ }\widetilde{X}^{(n)}(x) &=&n\int_{o}^{x}\widetilde{X}^{(n-1)}(\rho )%
\left[ f^{2}(\rho )\right] ^{(-1)^{n-1}}d\rho
\end{eqnarray*}%
and%
\begin{equation}
\psi _{k}(y)=\QATOPD\{ . {g(y)Y^{(k)}(y)\text{, }k\text{ odd}}{g(y)%
\widetilde{Y}^{(k)}(y)\text{, }k\text{ even}}  \label{10}
\end{equation}%
where%
\begin{eqnarray*}
Y^{(0)}(y) &=&\widetilde{Y}^{(0)}(y)=1 \\
\text{\ }Y^{(n)}(y) &=&n\int_{o}^{x}Y^{(n-1)}(\xi )\left[ g^{2}(\xi )\right]
^{(-1)^{n}}d\xi \\
\text{\ }\widetilde{Y}^{(n)}(y) &=&n\int_{o}^{x}\widetilde{Y}^{(n-1)}(\xi )%
\left[ g^{2}(\xi )\right] ^{(-1)^{n-1}}d\xi .
\end{eqnarray*}

Then the formal powers based on the given generating sequence are defined by
the formulas%
\begin{equation}
Z^{(n)}(\alpha ,0;z)=\phi (x,y)Sc_{\ast }Z^{(n)}(\alpha ,0;z)+\frac{\mathbf{k%
}}{\phi (x,y)}Vec_{\ast }Z^{(n)}(\alpha ,0;z)  \label{Zn}
\end{equation}%
where%
\begin{equation}
_{\ast }Z^{(n)}(\alpha ,0;z)=\QATOPD\{ . {\alpha 
{\acute{}}%
\dsum\limits_{m=0}^{n}\binom{n}{m}X^{(n-m)}\mathbf{k}^{m}\widetilde{Y}^{(m)}+%
\mathbf{k}\alpha 
{\acute{}}%
{\acute{}}%
\dsum\limits_{m=0}^{n}\binom{n}{m}\widetilde{X}^{(n-m)}\mathbf{k}^{m}Y^{(m)}%
\text{, }n\text{ odd}}{\alpha 
{\acute{}}%
\dsum\limits_{m=0}^{n}\binom{n}{m}\widetilde{X}^{(n-m)}\mathbf{k}^{m}%
\widetilde{Y}^{(m)}+\mathbf{k}\alpha 
{\acute{}}%
{\acute{}}%
\dsum\limits_{m=0}^{n}\binom{n}{m}X^{(n-m)}\mathbf{k}^{m}Y^{(m)}\text{, }n%
\text{ even.}}  \label{Znstar}
\end{equation}

In a similar way the formal powers corresponding to $\left( \ref{6}\right) $
can be constructed, they will be denoted as $\overset{\sim }{Z}^{(n)}(\alpha
,0;z)$. Notice that a generating pair for $\left( \ref{6}\right) $ is given
by 
\begin{equation*}
\overset{\sim }{F}_{0}=\frac{g}{f}\text{ \ \ and \ \ }\overset{\sim }{G}_{0}=%
\mathbf{k}\frac{f}{g}.
\end{equation*}

In \cite{CKT} it was shown that for functions $f$ and $g$ satisfying the
above conditions there exist the transmutation operators $T_{f}$ and $T_{g}$
defined as follows%
\begin{equation}
T_{f}\left[ u(x)\right] =u(x)+\int_{-x}^{x}\mathbf{K}(x,t;f%
{\acute{}}%
(0)).u(t)dt  \label{Tf}
\end{equation}%
where the kernel $\mathbf{K}(x,t;f%
{\acute{}}%
(0))$ is given by 
\begin{equation*}
\mathbf{K}(x,t;f%
{\acute{}}%
(0))=\frac{f%
{\acute{}}%
(0)}{2}+K(x,t)+\frac{f%
{\acute{}}%
(0)}{2}\int_{t}^{x}\left[ K(x,s)-K(x,-s)\right] ds
\end{equation*}%
and the function $K(x,t)$ is the unique solution of the Goursat problem (see 
\cite{Marchenko})%
\begin{equation*}
\QATOPD\{ . {\left( \frac{\partial ^{2}}{\partial x^{2}}-q(x)\right) K(x,t)=%
\frac{\partial ^{2}}{\partial t^{2}}K(x,t)}{K(x,x)=\frac{1}{2}%
\int_{0}^{x}q(s)ds\text{; }K(x,-x)=0}
\end{equation*}%
and%
\begin{equation*}
T_{g}\left[ v(y)\right] =v(y)+\int_{-y}^{y}\widetilde{\mathbf{K}}(y,t;g%
{\acute{}}%
(0))v(t)dt
\end{equation*}%
where%
\begin{equation*}
\widetilde{\mathbf{K}}(y,t;g%
{\acute{}}%
(0))=\frac{g%
{\acute{}}%
(0)}{2}+\widetilde{K}(y,t)+\frac{g%
{\acute{}}%
(0)}{2}\int_{t}^{y}\left[ \widetilde{K}(y,s)-\widetilde{K}(y,-s)\right] ds
\end{equation*}%
and the function $\widetilde{K}(x,t)$ is the unique solution of the Goursat
problem%
\begin{equation*}
\QATOPD\{ . {\left( \frac{\partial ^{2}}{\partial y^{2}}-\widetilde{q}%
(y)\right) \widetilde{K}(y,t)=\frac{\partial ^{2}}{\partial t^{2}}\widetilde{%
K}(y,t)}{\widetilde{K}(y,y)=\frac{1}{2}\int_{0}^{y}\widetilde{q}(s)ds\text{;
\ }\widetilde{K}(y,-y)=0}
\end{equation*}%
with $q=f\,^{\prime \prime }/f$ \ and \ $\widetilde{q}=g\,^{\prime \prime
}/g $.

Moreover, $T_{f}$ and $T_{g}$ satisfy the relations 
\begin{equation}
T_{f}\left[ x^{k}\right] =\varphi _{k}\ \text{\ \ and \ }\ T_{g}\left[ y^{k}%
\right] =\psi _{k},\ \forall k\in 
\mathbb{N}
_{0}.  \label{mapping powers 1}
\end{equation}%
We will need similar systems of functions $\left\{ \widetilde{\varphi }%
_{k}\right\} _{k=0}^{\infty }$ and $\left\{ \widetilde{\psi }_{k}\right\}
_{k=0}^{\infty }$ corresponding to $1/f$ and $1/g$ respectively, 
\begin{equation}
\widetilde{\varphi }_{k}(x)=\QATOPD\{ . {\frac{1}{f(x)}X^{(k)}(x)\text{, }k%
\text{ even}}{\frac{1}{f(x)}\widetilde{X}^{(k)}(x)\text{, }k\text{ odd}}
\label{11}
\end{equation}%
\begin{equation}
\widetilde{\psi }_{k}(y)=\QATOPD\{ . {\frac{1}{g(y)}Y^{(k)}(y)\text{, }k%
\text{ even}}{\frac{1}{g(y)}\widetilde{Y}^{(k)}(y)\text{, }k\text{ odd.}}
\label{12}
\end{equation}

For these systems of functions another pair of transmutations $T_{1/f}$ and $%
T_{1/g}$ is constructed (see \cite{KT}), one of the representations of which
can be given by the equalities 
\begin{equation*}
T_{1/f}u(x)=\frac{1}{f(x)}\left\{ \int_{0}^{x}f(\eta )T_{f}\left[ \partial
u(\eta )\right] d\eta +u(0)\right\} ,
\end{equation*}%
\begin{equation*}
T_{1/g}v(y)=\frac{1}{g(y)}\left\{ \int_{0}^{y}g(\eta )T_{g}\left[ \partial
v(\eta )\right] d\eta +v(0)\right\} .
\end{equation*}%
They satisfy the equalities 
\begin{equation}
T_{1/f}\left[ x^{k}\right] =\widetilde{\varphi }_{k}\ \text{and\ }T_{1/g}%
\left[ y^{k}\right] =\widetilde{\psi }_{k},\ \forall k\in 
\mathbb{N}
_{0}.  \label{mapping powers 2}
\end{equation}

The operators $T_{1/f}$ and $T_{1/g}$ admit the representations as Volterra
integral operators \cite{KT}, 
\begin{equation*}
T_{1/f}u(x)=u(x)+\int_{-x}^{x}\mathbf{K}_{2}(x,t;-f%
{\acute{}}%
(0))u(t)\,dt,
\end{equation*}%
where the kernel $\mathbf{K}_{2}(x,t;-f%
{\acute{}}%
(0))$ has the form 
\begin{equation*}
\mathbf{K}_{2}(x,t;-f%
{\acute{}}%
(0))=-\frac{1}{f(x)}\bigg(\int_{-t}^{x}\partial _{t}\mathbf{K}_{1}(s,t;f%
{\acute{}}%
(0))f(s)\,ds+\frac{f%
{\acute{}}%
(0)}{2}f(-t)\bigg)
\end{equation*}%
and the formulas for $T_{1/g}$ are completely analogous with an obvious
substitution of $f$ by $g$.

The introduced transmutation operators satisfy interesting commutation
equalities.

\begin{corollary}
\label{CorCommutation}\cite{KT} The following operator equalities hold on $%
C^{1}$-functions of the respective variables 
\begin{align}
\partial _{x}fT_{1/f}& =fT_{f}\partial _{x},\qquad \partial _{x}\frac{1}{f}%
T_{f}=\frac{1}{f}T_{1/f}\partial _{x}.  \label{CommutT1dx} \\
\partial _{y}gT_{1/g}& =gT_{g}\partial _{y},\qquad \partial _{y}\frac{1}{g}%
T_{g}=\frac{1}{g}T_{1/g}\partial _{y}.  \label{CommutT2dx}
\end{align}
\end{corollary}

Consider the operators projecting onto the scalar and the vector parts
respectively 
\begin{equation*}
P^{+}=\frac{1}{2}\left( I+C\right) \text{ \ and \ }P^{-}=\frac{1}{2\mathbf{k}%
}\left( I-C\right) .
\end{equation*}%
Let us introduce the following operators 
\begin{equation}
\mathbf{T}_{\mathbf{0}}\mathbf{=}T_{f}T_{g}P^{+}+\mathbf{k}%
T_{1/f}T_{1/g}P^{-}  \label{20}
\end{equation}%
and 
\begin{equation}
\mathbf{T}_{\mathbf{1}}\mathbf{=}T_{1/f}T_{g}P^{+}+\mathbf{k}%
T_{f}T_{1/g}P^{-}.  \label{41}
\end{equation}

From now on let $\Omega \subset \overline{R}=\left[ -a,a\right] \times \left[
-b,b\right] $ be a simply connected domain such that together with any point 
$(x,y)$ belonging to $\Omega $ the rectangle with the vertices $(x,y)$, $%
(-x,y)$, $(x,-y)$ and $(-x,-y)$ also belongs to $\Omega $. In such a domain
application of operators $\mathbf{T}_{\mathbf{0}}$ and $\mathbf{T}_{\mathbf{1%
}}$ is meaningful.

\begin{proposition}
\label{PropCommutationT}The following equalities hold for any $\mathbb{B}$%
-valued, continuously differentiable function $w$ defined in $\Omega $. 
\begin{equation}
\left( \overline{\partial }-\frac{\overline{\partial }\phi }{\phi }C\right) 
\mathbf{T}_{\mathbf{0}}w=\mathbf{T}_{\mathbf{1}}\left( \overline{\partial }%
w\right) ,\qquad \left( \overline{\partial }+\frac{\partial \phi }{\phi }%
C\right) \mathbf{T}_{\mathbf{1}}w=\mathbf{T}_{\mathbf{0}}\left( \overline{%
\partial }w\right) .  \label{PropertiesT}
\end{equation}%
\begin{equation}
\left( \partial -\frac{\partial \phi }{\phi }C\right) \mathbf{T}_{\mathbf{0}%
}w=\mathbf{T}_{\mathbf{1}}\left( \partial w\right) ,\qquad \left( \partial +%
\frac{\overline{\partial }\phi }{\phi }C\right) \mathbf{T}_{\mathbf{1}}w=%
\mathbf{T}_{\mathbf{0}}\left( \partial w\right) .  \label{PropertiesTder}
\end{equation}
\end{proposition}

\begin{proof}
The proof consists in a direct calculation with the aid of the relations
from Corollary \ref{CorCommutation}.
\end{proof}

An immediate corollary of equalities (\ref{PropertiesT}) is the fact that
the operator $\mathbf{T}_{\mathbf{0}}$ maps bicomplex analytic functions
into $\left( \phi ,\mathbf{k}/\phi \right) -$pseudoanalytic, i.e., into
solutions of $\left( \ref{8}\right) $ and the operator $\mathbf{T}_{\mathbf{1%
}}$ maps bicomplex analytic functions into $\left( \frac{g}{f},\mathbf{k}%
\frac{f}{g}\right) $-pseudoanalytic i.e., into solutions of the equation 
\begin{equation}
\left( \overline{\partial }+\frac{\partial \phi }{\phi }C\right) W=0.
\label{VekuaSucceed}
\end{equation}
Moreover, they map powers of the variable $z$ into corresponding formal
powers.

\begin{proposition}
\label{PropMapComplexPowers}For any $z\in \Omega $ and $a\in \mathbb{B}$ the
following equalities are valid 
\begin{equation*}
\mathbf{T}_{\mathbf{0}}[az^{n}]=Z^{(n)}(a,0;z)\quad \text{and}\quad \mathbf{T%
}_{\mathbf{1}}[az^{n}]=Z_{1}^{(n)}(a,0;z).
\end{equation*}
\end{proposition}

\begin{proof}
The proof consists in the observation that for $a=a%
{\acute{}}%
+\mathbf{k}b%
{\acute{}}%
$ \ and \ $z=x+\mathbf{k}y$ one has%
\begin{equation*}
az^{n}=\left( a%
{\acute{}}%
+\mathbf{k}b%
{\acute{}}%
\right) \dsum\limits_{m=0}^{n}\binom{n}{m}x^{n-m}\mathbf{k}^{m}y^{m}
\end{equation*}%
and the result follows from the formulas (\ref{Zn}), (\ref{Znstar}) by
application of the mapping properties (\ref{mapping powers 1}), (\ref%
{mapping powers 2}).
\end{proof}

Notice that both $\mathbf{T}_{\mathbf{0}}$ and $\mathbf{T}_{\mathbf{1}}$ are
bounded operators on the space of continuous functions with respect to the
norm $\left\Vert w\right\Vert =\max (\left\vert u\right\vert +\left\vert
v\right\vert )$ where $w=u+\mathbf{k}v$. Indeed, consider $\left\Vert 
\mathbf{T}_{\mathbf{0}}w\right\Vert =\max (\left\vert T_{f}T_{g}u\right\vert
+\left\vert T_{1/f}T_{1/g}v\right\vert )\leq M_{1}\max \left\vert
u\right\vert +M_{2}\max \left\vert v\right\vert $ where the constants $M_{1}$
and $M_{2}$ depend only on the corresponding kernels of the bounded Volterra
operators $T_{f}$, $T_{g}$, $T_{1/f}$ and $T_{1/g}$. Then $\left\Vert 
\mathbf{T}_{\mathbf{0}}w\right\Vert \leq M\left\Vert w\right\Vert $ where $%
M=\max \left\{ M_{1},M_{2}\right\} $. The proof for the operator $\mathbf{T}%
_{\mathbf{1}}$ is analogous. Moreover, $\mathbf{T}_{\mathbf{0}}^{-1}$ and $%
\mathbf{T}_{\mathbf{1}}^{-1}$ are bounded as well (the form of the inverses
for $T_{f}$, $T_{g}$, $T_{1/f}$ and $T_{1/g}$ can be found in \cite{KT}) due
to the fact that their integral kernels enjoy the same regularity properties
as the kernels of $\mathbf{T}_{\mathbf{0}}$ and $\mathbf{T}_{\mathbf{1}}$.

Let us establish another useful fact concerning the mapping properties of
the operators $\mathbf{T}_{\mathbf{0}}$ and $\mathbf{T}_{\mathbf{1}}$.

\begin{proposition}
\label{PropDerivatives}Let $w$ be a bicomplex analytic function in $\Omega $
and $W=\mathbf{T}_{\mathbf{0}}w$ be a corresponding solution of $\left( \ref%
{8}\right) $. Then 
\begin{equation}
\mathbf{T}_{\mathbf{0}}\left( \partial ^{(2n)}w\right) =W^{\left[ 2n\right]
}\quad \text{and\quad }\mathbf{T}_{\mathbf{1}}\left( \partial
^{(2n-1)}w\right) =W^{\left[ 2n-1\right] }\text{,\quad }n=1,2,\ldots .
\label{Relations for derivatives}
\end{equation}
\end{proposition}

\begin{proof}
From $\left( \ref{PropertiesTder}\right) $ we have%
\begin{equation}
\overset{.}{W}=\mathbf{T}_{\mathbf{1}}\left( \partial w\right) .  \label{45}
\end{equation}%
$\overset{.}{W}$ is a solution of the succeeding Vekua equation (\ref%
{VekuaSucceed}). Denote $W_{1}=\overset{.}{W}$. Any solution of (\ref%
{VekuaSucceed}) is the image of a bicomplex analytic function under the
action of the operator $\mathbf{T}_{\mathbf{1}}$, so $W_{1}=\mathbf{T}_{%
\mathbf{1}}w_{1}$. Due to $\left( \ref{PropertiesTder}\right) $ we have $%
\overset{.}{W_{1}}=\mathbf{T}_{\mathbf{0}}\left( \partial w_{1}\right) $.
Thus, $\overset{..}{W}=\mathbf{T}_{\mathbf{0}}\left( \partial ^{2}w\right) $
because from (\ref{45}) $w_{1}=\partial w$. Now (\ref{Relations for
derivatives}) can be easily proved by induction.
\end{proof}

The established relations from Propositions \ref{PropCommutationT} and \ref%
{PropMapComplexPowers} together with the fact that $\mathbf{T}_{\mathbf{0}}$
and $\mathbf{T}_{\mathbf{1}}$ are bounded operators together with their
respective inverses allow us to transfer several results from analytic
function theory onto the solutions of the bicomplex Vekua equations under
consideration and as hence onto the solutions of the Dirac system with a
scalar potential being a function of one Cartesian variable. Here we give
two examples of such results.

\begin{theorem}
Let $W$ be a solution of $\left( \ref{8}\right) $ in a disk $D$ with the
center in the origin and radius $R$. Then it can be expanded into a Taylor
series in formal powers 
\begin{equation*}
W(z)=\dsum\limits_{n=0}^{\infty }Z^{(n)}(a_{n},0;z)
\end{equation*}%
with the radius of convergence $R$. The series converges normally in $D$ and
the coefficients $a_{n}$ have the form 
\begin{equation*}
a_{n}=\frac{W^{\left[ n\right] }(0)}{n!}.
\end{equation*}

\begin{proof}
Consider $w=\mathbf{T}_{\mathbf{0}}^{-1}W$. It is a bicomplex analytic
function, so we have that it can be expanded into a Taylor series $%
w(z)=\dsum\limits_{n=0}^{\infty }a_{n}z^{n}$ with the coefficients $a_{n}=%
\frac{d^{n}w(0)}{dz^{n}}/n!$. Application of $\mathbf{T}_{\mathbf{0}}$ gives
us a series for $W$, $W(z)=\mathbf{T}_{\mathbf{0}}w(z)=\dsum\limits_{n=0}^{%
\infty }\mathbf{T}_{\mathbf{0}}[a_{n}z^{n}]=\dsum\limits_{n=0}^{\infty
}Z^{(n)}(a_{n},0;z)$. Due to the uniform boundedness of $\mathbf{T}_{\mathbf{%
0}}$ the radius of convergence of the series is preserved. Note that the
Taylor coefficients coincide. In order to finish the proof we use
Proposition \ref{PropDerivatives} and the fact that both operators $\mathbf{T%
}_{\mathbf{0}}$ and $\mathbf{T}_{\mathbf{1}}$ preserve the values of a
function in the origin. This is obvious from their definition and from the
Volterra integral form of the operators $T_{f}$, $T_{g}$, $T_{1/f}$ and $%
T_{1/g}$ (see, e.g., (\ref{Tf})). Thus, $W^{\left[ 2n\right] }(0)=\mathbf{T}%
_{\mathbf{0}}\left( \partial ^{(2n)}w\right) (0)=\partial ^{(2n)}w(0)$ and $%
W^{\left[ 2n-1\right] }(0)=\mathbf{T}_{\mathbf{1}}\left( \partial
^{(2n-1)}w\right) (0)=\partial ^{(2n-1)}w(0)$,\quad $n=1,2,\ldots $.
\end{proof}
\end{theorem}

\begin{theorem}
Any solution $W$ of $\left( \ref{8}\right) $ in $\Omega $ can be \
approximated arbitrarily closely on any compact subset $K$ of $\Omega $ by a
finite combination of formal powers (a formal polynomial) $%
\dsum\limits_{n=0}^{N}Z^{(n)}(a_{n},0;z)$.
\end{theorem}

\begin{proof}
Consider $w=\mathbf{T}_{\mathbf{0}}^{-1}W$. Due to the Runge approximation
theorem the function $w$ can be arbitrarily closely approximated by a
polynomial in $z$. Then due to the boundedness of $\mathbf{T}_{\mathbf{0}}$
and $\mathbf{T}_{\mathbf{0}}^{-1}$ and Proposition \ref{PropMapComplexPowers}
we obtain the required result for $W$.
\end{proof}

This theorem in fact means the completeness of the family of functions 
\begin{equation*}
\left\{ Z^{(n)}(1,0;z)\text{, \ }Z^{(n)}(\mathbf{k},0;z)\right\}
_{n=0}^{\infty }
\end{equation*}%
in the space of all solutions of the Vekua equation (\ref{8}). A similar
fact is true for equation (\ref{VekuaSucceed}) and corresponding formal
powers. The combination of both families of formal powers gives us a
complete family of solutions of the Dirac equation (\ref{ROmega}) in the
considered case.

\end{document}